\documentclass[conference,a4paper]{IEEEtran}
\usepackage{balance}
\usepackage[section]{placeins}
\usepackage{mwe}
\usepackage{lipsum}
\usepackage{adjustbox}
\usepackage{cite}
\usepackage{amsmath,amssymb,amsfonts}
\usepackage{algorithmic}
\usepackage{graphicx,color}
\usepackage{textcomp}
\usepackage{subcaption}
\usepackage[nolist,nohyperlinks]{acronym}
\usepackage{multirow}
\usepackage{amsmath}
\usepackage[linesnumbered,ruled,vlined]{algorithm2e}
\SetKw{KWInitiate}{Initiate:}
\SetKw{KWCalculate}{Calculate:}
\SetKw{KWGenerate}{Generate:}
\SetKw{KWOptimize}{Optimize:}
\SetKw{KWExtract}{Extract:}
\SetKw{KWCrossOver}{CrossOver:}
\SetKw{KWMutate}{Mutate:}
\usepackage[table,xcdraw]{xcolor}	
\SetKwInput{KwInput}{Input}                % Set the Input
\SetKwInput{KwOutput}{Output}              % set the Output

\begin{acronym}
\acro{DRA}{Direct Radiating Array}
\acro{FoV}{Field of View}
\acro{SLL}{Side Lobe Level}
\acro{EIRP}{Effective Isotropic Radiated Power}
\acro{LEO}{Low Earth Orbit}
\acro{MVDR}{Minimum Variance Distortionless Response}
\acro{LCMV}{Linearly Constrained Minimum Variance}
\acro{GA}{Genetic Algorithm}
\acro{NLEO}{Not Low Earth Orbit}
\acro{GEO}{Geostationary Orbit}
\acro{NGSO}{Non-Geostationary Orbit}
\acro{LHCP}{Left-Hand Circular Polarization}
\acro{RHCP}{Right-Hand Circular Polarization}
\acro{RF}{Radio Frequency}
\acro{ADC}{Analog-to-Digital Converter}
\acro{DAC}{Digital-to-Analog Converter}
\acro{LNA}{Low-noise Amplifier}
\acro{FFT}{Fast Fourier Transform}
\acro{ZF}{Zero Forcing}
\acro{SVD}{Singular Value Decomposition}

\end{acronym}
\ifCLASSOPTIONcompsoc
 \usepackage[caption=false,font=normalsize,labelfont=sf,textfont=sf]{subfig}
\else
 \usepackage[caption=false,font=footnotesize]{subfig}
\fi
\begin{document}
%
% paper title
% Titles are generally capitalized except for words such as a, an, and, as,
% at, but, by, for, in, nor, of, on, or, the, to and up, which are usually
% not capitalized unless they are the first or last word of the title.
% Linebreaks \\ can be used within to get better formatting as desired.
% Do not put math or special symbols in the title.
\title{ Optimizing Antenna Activation for Even Power Distribution in Multi-Beam Satellite Systems Using Genetic Algorithm}

% author names and affiliations
% use a multiple column layout for up to three different
% affiliations
\author{\IEEEauthorblockN{
Juan Andr\'es V\'asquez-Peralvo, Aral Ertug Zorkun, Jorge Querol, Eva Lagunas, Flor Ortiz,\\ 
Luis Manuel Garcés-Socarrás, Jorge Luis Gonz\'alez-Rios, Symeon Chatzinotas.     
}                                     % ...
%\\
\IEEEauthorblockA{% 1st affiliations
Interdisciplinary Centre for Security Reliability and Trust, University of Luxembourg, 1855 Luxembourg-Luxembourg\\ (e-mails: \{ juan.vasquez, aral.zorkun, jorge.querol, eva.lagunas, flor.ortiz, luis.garces, \\jorge.gonzalez, victor.monzon, symeon.chatzinotas\}@uni.lu).}

}

% conference papers do not typically use \thanks and this command
% is locked out in conference mode. If really needed, such as for
% the acknowledgment of grants, issue a \IEEEoverridecommandlockouts
% after \documentclass

% use for special paper notices
%\IEEEspecialpapernotice{(Invited Paper)}

% make the title area
\maketitle

% As a general rule, do not put math, special symbols or citations
% in the abstract
\begin{abstract}
Recent advancements in onboard satellite communication have significantly enhanced the ability to dynamically modify the radiation pattern of a \ac{DRA}, which is essential for both conventional communication satellites like \ac{GEO} and those in lower orbits such as \ac{LEO}. This is particularly relevant for communication at 28 GHz, a key frequency in the mmWave spectrum, used for high-bandwidth satellite links and 5G communications. Critical design factors include the number of beams, beamwidth, and \ac{SLL} for each beam. However, in multibeam scenarios, balancing these design factors can result in uneven power distribution, leading to over-saturation in centrally located antenna elements due to frequent activations. This paper introduces a \ac{GA}-based approach to optimize beamforming coefficients by modulating the amplitude component of the weight matrix, while imposing a constraint on activation instances per element to avoid over-saturation in the \ac{RF} chain. The proposed method, tested on an 16×16 \ac{DRA} patch antenna array at 28 GHz for a CubeSat orbiting at 500 km, demonstrates how the algorithm efficiently meets beam pattern requirements and ensures uniform activation distribution. These findings are particularly relevant for emerging satellite systems and 5G networks operating in the mmWave spectrum.
\end{abstract}

\vskip0.5\baselineskip

% For peer review papers, you can put extra information on the cover
% page as needed:
% \ifCLASSOPTIONpeerreview
% \begin{center} \bfseries EDICS Category: 3-BBND \end{center}
% \fi
%
% For peerreview papers, this IEEEtran command inserts a page break and
% creates the second title. It will be ignored for other modes.
% \IEEEpeerreviewmaketitle

\section{Introduction}

 Digital beamforming is becoming easier to get and use, which is making phased arrays even more popular \cite{brookner2002phased}. Since there is a trade-off between signal processing and hardware, new applications are emerging where the phased antenna arrays are in need \cite{herd2015evolution}. However, cost and power consumption push the digital beamforming into the background
 \cite{vasquez2024multibeam} where each antenna element requires active \ac{RF} components \cite{palisetty2023fpga}. Therefore, power optimization is crucial, especially in satellite systems, where flexible payloads are the most important \cite{ortiz2024harnessing}.

In the context of flexible beamforming within a phased array system, achieving the desired beam characteristics — such as adjustable beam shape, \ac{SLL}, and null steering is essential for adaptability. Deterministic beamforming techniques, such as \ac{FFT}, \ac{ZF}, and \ac{SVD}, can be employed to steer beams in specific directions while creating nulls in others. \ac{SLL} control can be achieved through amplitude tapering, where the power distribution across antenna elements is varied \cite{mailloux2018phased}. However, these methods are not sufficient for generating complex beam shapes nor suitable for applying amplitude control to achieve specific beamwidths or sidelobe levels where power efficiency and heat dissipation are critical (multibeam scenarios) \cite{vasquez2023flexible}. Another approach is the use of artificial intelligence to determine the required weight vector to address required beam characteristics with an approach of optimized power consumption \cite{fontanesi2023artificial}, but his approach requires a data set that must be obtained using any other technique.

To address these challenges, this paper proposes a solution for generating beams with flexible beamwidth, and \ac{SLL} by implementing array thinning to avoid reliance on amplitude tapering. By reducing the number of active elements in a controlled manner, array thinning helps maintain power efficiency while achieving the desired beam characteristics. Additionally, we introduce a constraint to distribute power more evenly across antenna elements by limiting their activation instances. In a typical scenario, generating multiple beams would require each antenna element to deliver power for all beams simultaneously, resulting in excessive power demands. In contrast, our proposed approach limits the number of antenna activation per beam. Once the maximum activation limit is reached for a given element, the algorithm dynamically selects a free element to take over, ensuring an even distribution of power and reducing thermal stress across the array.

\section{Radiation Pattern Parameters}
This section describes the antenna that will be used, its parameters that will be considered in the optimization algorithm. 
\subsection{Unit Cell}
The antenna used in this analysis is an open-ended waveguide since we are targeting low losses and high directivity. The radiation pattern of the open-ended waveguide can be modeled using the following electric field $ \mathbf{E}(r, \theta, \phi)$ formula:

\begin{equation}
\mathbf{\hat{E}}(r, \theta, \phi) = E_0 \frac{J_1\left(k a \sin\theta\right)}{k a \sin\theta} \frac{e^{-jkr}}{r} \left( \cos(\phi) \hat{\theta} + j \sin(\phi) \hat{\phi} \right)
\label{eq:Equation9}
\end{equation}

where:
\begin{itemize}
    \item $ E_0 $ is the amplitude scaling factor.
    \item $ J_1 $ represents the Bessel function of the first kind of order 1.
    \item $ k = \frac{2\pi}{\lambda} $ is the wave number.
    \item $ a $ denotes the radius of the waveguide.
    \item $ r $ indicates the distance from the waveguide's aperture.
    \item $ \theta $ and $ \phi $ are the polar and azimuthal angles, respectively.
    \item $ \hat{\theta} $ and $ \hat{\phi} $ are unit vectors in the polar and azimuthal directions.
\end{itemize}

For easiness of computation and understanding of resules, we compute the normalized power pattern as \cite{vasquez2021interwoven}:
\[
{\mathbf{U}}(\theta, \phi) = \frac{\left| E(\theta, \phi) \right|^2}{\text{max}\left(\left| E(\theta, \phi) \right|^2\right)}
\]

In this paper we select an antenna aperture of 0.58 $\lambda$ (a = 0.3$\lambda$) and an inter element space between antennas of 0.6$\lambda$.
\subsection{Antenna Parameters}
The parameters included in the optimization process are: beamwidth and \ac{SLL}. Following, we will describe each of them and how can be extracted from the calculated radiation pattern.

\subsubsection{Beamwidth}
The average value obtained from different radiation pattern cuts is used to determine the beamwidth of each beam in the radiation pattern. The angular difference between the highest and minimum angles at which the -3dB beamwidth point occurs is used to calculate the beamwidth for each cut. This calculation method is specified in \eqref{eq:beamwidth}.

\begin{equation}
    \label{eq:beamwidth}
    \theta_{-3dB}^b = |\Theta_{-3_{dB, 2}}^b - \Theta_{-3_{dB, 1}}^b|,
\end{equation}
where $\Theta_{-3_{dB, 1}}$, $\Theta_{-3_{dB, 2}}$ are the angles at which the radiated power drops to half.

\subsubsection{Side Lobe Levels}
The \ac{SLL}, which is measured in decibels (dB), is the ratio of the side lobes' maximum amplitude to the main lobe's maximum amplitude. as shown in \eqref{eq:sll}:

\begin{equation}
\label{eq:sll}
\text{SLL}^b \text{(dB)} = 20 \log_{10} \left( \frac{\text{AP}_{\text{SL,max}}^b}{\text{AP}_{\text{max}}^b} \right),
\end{equation}

The \ac{SLL} might not always be in a principal cut throughout this optimization process, which entails turning on and off antenna elements. Instead, it could show up at any arbitrary angular point. In order to determine the overall \ac{SLL}, the optimizer will look at the complete radiation pattern and find the second-highest maximum, omitting those that are outside the field of view.

\section{Array Thinning Algorithm}
In large array structures with hundreds of antennas, array thinning is usually done by turning off a certain number of antenna elements depending on the desired \ac{SLL}. The main benefit of multi-beam array thinning is its contribution to the optimization process in terms of \ac{SLL} and beamwidth. Since the number of antenna elements is quite large additional flexibility and degrees of freedom can be achieved for the desired results. Therefore, the proposed algorithm first takes into account the desired $\theta_{-3dB}$, and \ac{SLL}. 

The minimization problem for the calculated beamwidth ($\mathrm{\theta_{-3dB_{c}} }(W_{m\times n} )$), and minimum Side Lobe Level ($\mathrm{SLL_{c} }(W_{m\times n} )$) with the desired counterparts ($\mathrm{\theta_{-3dB_{Az_o}} }$), ($\mathrm{SLL_{o} }(W_{m\times n} )$), respectively, can be defined as:

\begin{equation}
\label{eq:CostFunction1}
%\begin{split}
\min_{\vspace{1cm} W_{m\times n} }     \hspace{1cm}             Z_1(W_{m\times n} )+Z_2(W_{m\times n} )
%\end{split}
\end{equation}
where:

\begin{equation*}
\label{eq:Eq2bF}
    \left\{
    \begin{aligned}
        & Z_1 =  \sum_{r=1}^R\Bigg( \frac{|\mathrm{ \theta_{-3dB_{r}} }(W_{m\times n} )-\mathrm{\theta_{-3dB_{o}} }|}{\mathrm{\theta_{-3dB_{o}} }}\Bigg)k_1 \\ 
        & Z_2 = %\Bigg(\frac{|\mathrm{SLL_{c} }(W_{m\times n} )-\mathrm{SLL_{o} }|}{\mathrm{SLL{Az_o} }}\Bigg)k_2\\
        \begin{cases}
        0, & \text{if } SLL_c(W_{m\times n}) > 16 \quad \\
         \frac{|SLL_c(W_{m\times n})-SLL_0|}{SLL_0} & \text{if }  SLL_c(W_{m\times n})\leq 16,  \\
    \end{cases}
        \end{aligned}
    \right.
\end{equation*}
and $k_1$ and $k_2$ are weights that determine the relative importance of achieving the desired beamwidth versus the side lobe level, $m$ and $n$  are the number of antenna elements in $x$-axis and $y$-axis, respectively, and R is the number of cuts where the beamwidth will be assessed.
The goal of the proposed algorithm is to compute the weight vector, which is crucial for achieving the desired beamwidth and \ac{SLL}. The weights represent the amplitude and phase of each antenna element. The weight of the $mn$-th antenna element, $W_{m \times n}$, in a planar array can be calculated as:

\begin{align}
\label{eq:scanning_1}
    W_{mnp}=  |W_{mnp}|{\rm e}^{-{\rm j}\kappa (\sin\theta_0(m d_x \cos\phi_0+n d_y \sin\phi_0) + p d_z cos\theta_0 )},\
\end{align}
where ($\theta_0$, $\phi_0$) represent the scanning angles.

The possibility of an optimization procedure that, particularly when synthesizing beams with various constraints, identifies appropriate configurations to balance antenna element activation. Therefore, in multi-beam scenarios, the number of activation instances of each antenna element can be taken into account as an optimization parameter. 

By adopting evenly distributed activation throughout the array, this method improves the system's efficiency as well as overall performance. In addition, by optimizing heat management and power distribution, the sustainability and reliability of the system can be increased.

As a result of controlling activation instances, overall power consumption can be limited by introducing a total power constraint to the system while maintaining the desired beamwidth.  

The optimization problem for the multi-beam scenario can be mathematically represented as:

\begin{equation}
    \begin{aligned}
        \min_{\vspace{1cm} W^b } \quad & f(\theta_{-3dB}^b(W^b), SLL^b(W^b)) \\
        \text{subject to} \quad & \sum_{b=1}^B \sum_{m = 1}^M \sum_{n = 1}^N |W_{m,n}^b|^2\leq p_\text{max} \\
        & \sum_{b=1}^B |W^b|\odot |W^b| \leq P_\text{max}, \\
       % & x_{\text{lower}} \leq x \leq x_{\text{upper}}
    \end{aligned}
    \label{eq:functionMultibeam}
\end{equation}
where $f$ is the cost  function and it is defined by \eqref{eq:costFunctionMultibeam}:

    \begin{equation}
        \begin{aligned}
             \quad & f(\theta^b_{-3dB}, SLL^b)= Z_1(W^b) + Z_2(W^b) \\
           & Z_1(W^b) = \frac{1}{B} \sum_{b = 1}^{b} \frac{1}{R} \sum _{r =1}^R \frac{|\theta_{-3dB_{r}}^b(W^b)-\theta_{-3dB_{0}}^b(W^b)|}{\theta_{-3dB_{0}}^b(W^b)}\\
    &Z_2(W^b) =
    \begin{cases}
        0, & \text{if } \text{SLL}_c^b > 16 \quad \\
        \frac{1}{B} \sum_{b = 1}^{B}\frac{|\text{SLL}_c^b(W^b)-\text{SLL}_0|}{\text{SLL}_0}, & \text{if }  \text{SLL}_c^b\leq 16  \\
    \end{cases}
    \end{aligned}
    \label{eq:costFunctionMultibeam}
\end{equation}
where $\theta^b_{-3dB_{0}}$ is the desired -3 dB beamwidth per beam, $\theta^b_{-3dB_{c}}$ is the calculated -3 dB beamwidth per beam, $SLL^b_0$ is the overall desired SLL per beam, $SLL^b_c$ is the calculated Side Lobe level per beam, $W^b$ is the calculated weight matrix per beam, $ p_\text{max}$ is the maximum power of the array, and  $P_\text{max}$ is the maximum activation times per element matrix.

As a final remark, the proposed multi-beam scenario involves the superimpose of beam power layers which are created concurrently. When operating at the same frequency, superimposing phases require spatial separation to prevent beam distortion. In order to eliminate such a problem, the graph coloring is adopted where each beam is generated by using separate frequency sub-bands. Consequently, the intended beam qualities are preserved by preventing interference.

Following a genetic type of optimization, we derived Algorithm \ref{algm:Beamforming2} \cite{vasquez2024multibeam} to find the suitable weight matrices that satisfy the problem presented in \eqref{eq:functionMultibeam}.

\begin{algorithm}[!ht]
\DontPrintSemicolon
  
  \KwInput{$(\Lambda^b,\Phi^b)$, center of the beam in Latitude and longitude coordinates per beam,\\ \hspace{1cm}$\mathrm{\theta^b_{-3dB_o}}$, required beamwidth per beam,\\ 
  \hspace{1cm}$p_\mathrm{max}$, maximum power consumption,\\
  \hspace{1cm}$P_\mathrm{max}$, Maximum activation times matrix,\\
  \hspace{1cm}$\mathrm{SLL_{o}^b}$, minimum SLL per beam,\\ } 
  \KwOutput{$W^b$, Weight matrix based on previous inputs}
  \KwData{Set of possible configurations on Satellite considering system constraints}
  %$F_1=$ initial value\\
  \begin{algorithmic}[1]
  \ENSURE  $\sum_{b=1}^B |W^b|\odot |W^b| \leq P_\text{max}$
  \ENSURE $\sum_{b=1}^B \sum_{m = 1}^M \sum_{n = 1}^N |W_{m,n}^b|^2\leq p_{\mathrm{\text{max}}}$
  \STATE \KWInitiate{$\mathrm{Population}$($W_o^b$) }\\
   \WHILE{Termination criteria not met}
  \STATE \KWCalculate{$\mathrm{Antenna}$ $\mathrm{Pattern}$ $F(W_c^b)$}\\
  \STATE \KWExtract{$\mathrm{\theta^b_{-3dB_c}}$, $\mathrm{SLL_{c}^b}$,  $p_\mathrm{max}$ $\mathrm{and}$ $P_\mathrm{max}$ }\\
 \STATE \KWCalculate{$\mathrm{f}(Z_1(W_c^b) + Z_2(W_c^b) )$}\\
              \IF{$f<f_{\mathrm{min}}$}
                  
                  \STATE \textbf{Save:} Optimal matrix $W^b = W_{c}^b$\\
                 \STATE \textbf{break}
                  
                  \ELSE
                   \STATE{ \textbf{Select:} Chromosomes based on fitnes}\\
                  \STATE {\textbf{Operations:} CrossOver and Mutation}\\
                  \STATE {\textbf{Create:} New Generation $W_c^b$}\\
                  \ENDIF

          \ENDWHILE      
    \end{algorithmic}

\caption{Beamforming Algorithm.}\label{algm:Beamforming2}
\end{algorithm}

The steps of the proposed genetic-based beamforming algorithm are described below:

\begin{itemize}
    \item \textbf{Step 1:} Random initial solution is generated in a tensor where its dimensions are $M$x$B$x$N$x$C$, $N$, where $M$ is the number of rows, $N$ the number of columns, $B$ the number of beams, and $C$ the total number of chromosomes (referred as initial population $W_o^b$ ).
    \item \textbf{Step 2:} Divide the tensor in dimensions of $M$x$b$, $N$, $C$, where $b$ is the  beam number.
    \item \textbf{Step 3:} Sort matrices from the lowest to the highest calculated cost values.
    \item \textbf{Step 4:} Eliminate the less efficient half of the stack in Step 3.
    \item \textbf{Step 5:} Repeat the procedure until the threshold. The evolutionary optimization balances the beamwidth and SLL parameters.
\end{itemize}

\section{Simulation Results}
Given the general characteristics of our evaluation setup,
we have analyzed the efficiency of the proposed algorithm under three different scenarios in terms of activation instances. The common parameters for all of the scenarios are given below:
\begin{itemize}
    \item \textbf{Operating Frequency:} 28 GHz.
    \item \textbf{Antenna Array:} 16 x 16 lattice planar array.
    \item \textbf{Orbital Altitude of the satellite:} The orbit of the satellite is considered 500 km above the surface of Earth.
    \item \textbf{Available total power:} We consider 40 dBm total power.
    \item \textbf{SLL:} The SLL is set to 16 dB for each of the synthesized beams.
    \item \textbf{Desired beam widths:} The directions of the desired beams widths in terms of elevation in degrees are given as  [4.7$^\circ$, 5.5$^\circ$, 6$^\circ$, 6.5$^\circ$, 
    5.8$^\circ$, 5$^\circ$, 7$^\circ$].
\end{itemize}

\subsection{Scenario \#1}
In Scenario \#1, the beam synthesis process is subject to specific requirements, as outlined below:

\begin{itemize}
    \item \textbf{Activation instances:} A limit of a maximum of 5 activation instances is imposed on each antenna element.
\end{itemize}

\begin{figure} [!htbp]  
\centering
\includegraphics[width=0.8\columnwidth]
{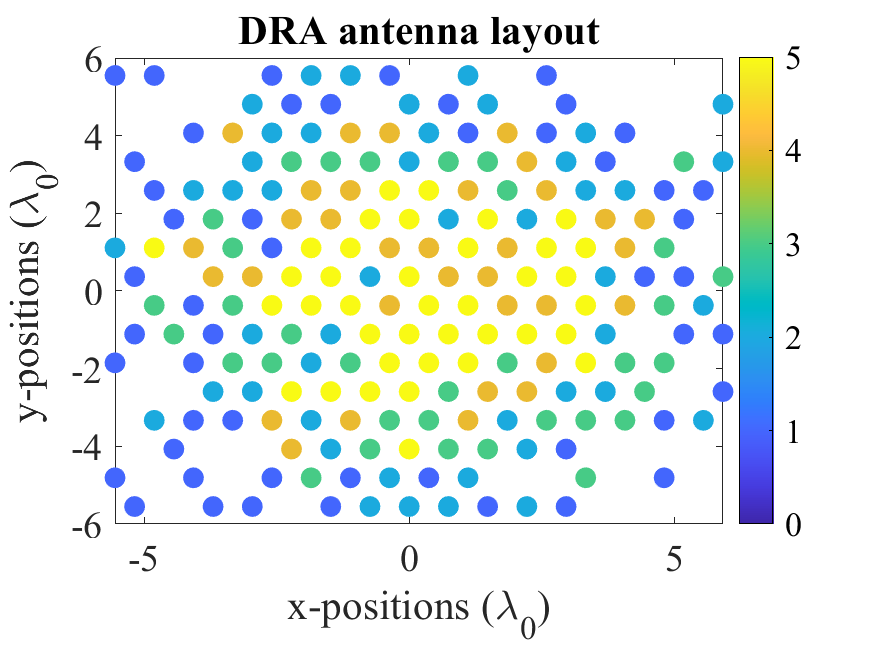}
\caption{Activation instances in the DRA for the Scenario \#1.}
\label{fig:Act5}
\end{figure}

Detailed results from the beam synthesis Scenario \#1 are presented in TABLE \ref{Table:Scenario1}. The activation instances for each antenna element in
Scenario \#1 is depicted in Fig. \ref{fig:Act5}

% Table 1
\begin{table}[!ht]
\caption{Characteristics of Synthesized Beams in the Scenario \#1.}
\label{Table:Scenario1}
\centering
\setlength{\tabcolsep}{0.25em}
\def\arraystretch{1.3}
\begin{tabular}{|c|c|c|c|c|c|c|}
%\begin{tabular}{|c|c|S|S|S|S|S|S|}
\hline
\textbf{Beam} & \textbf{\begin{tabular}[c]{@{}c@{}}$\theta_{-3dB_o}$\\ ($^\circ$)\end{tabular}} & \multicolumn{1}{c|}{\textbf{\begin{tabular}[c]{@{}c@{}}$\theta_{-3dB_c}$\\ ($^\circ$)\end{tabular}}} & \multicolumn{1}{c|}{\textbf{\begin{tabular}[c]{@{}c@{}}error\\ (\%)\end{tabular}}} & \multicolumn{1}{c|}{\textbf{\begin{tabular}[c]{@{}c@{}}SLL\\ (dB)\end{tabular}}} & \multicolumn{1}{c|}{\textbf{\begin{tabular}[c]{@{}c@{}}Act.\\ Elem.\end{tabular}}} & \textbf{\begin{tabular}[c]{@{}l@{}}Directivity\\ (dBi)\end{tabular}}  \\ \hline
\rowcolor[HTML]{FFFFFF} 
1  & 4.7 & 4.6827 & 0.3679 & 15.9 & 100 & 27.251 
\\ \hline
2  & 5.5  & 5.4812 & 0.3418 & 15.122 & 78 & 26.105 
\\ \hline                                                 
3  & 6 & 6.0658 & 1.0975 & 16.436  & 77 & 26.388  
\\ \hline
4 & 6.5 & 6.4505  & 0.7618 & 16.056 & 73 & 26.091 
\\ \hline                            5 & 5.8 & 5.8337 & 0.581 & 16.766 & 86 & 26.785 
\\ \hline             
6 & 5 & 4.9132 & 1.7363 & 14.51 & 87 & 26.548        \\ \hline                       
7 & 7 & 7.0224 & 0.3196 & 14.422 & 65 & 25.624   
\\ \hline                                                                                                            
\end{tabular}
\end{table}
% Table1 end

\subsection{Scenario \#2}
In Scenario \#2, the beam synthesis process is subject to specific requirements, as outlined below:

\begin{itemize}
    \item \textbf{Activation instances:} A limit of a maximum of 6 activation instances is imposed on each antenna element.
\end{itemize}

\begin{figure} [!htbp]  
\centering
\includegraphics[width=0.8\columnwidth]
{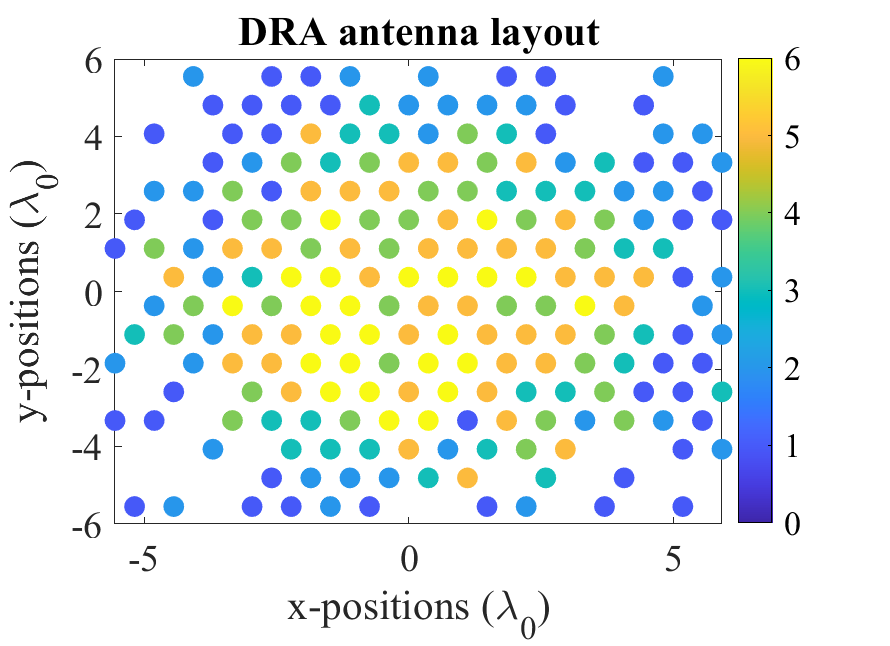}
\caption{Activation instances in the DRA for the Scenario \#2.}
\label{fig:Act6}
\end{figure}

Detailed results from the beam synthesis Scenario \#2 are presented in TABLE \ref{Table:Scenario2}. The activation instances for each antenna element in
Scenario \#2 is depicted in Fig. \ref{fig:Act6}

% Table 3
\begin{table}[!ht]
\caption{Characteristics of Synthesized Beams in the Scenario \#2.}
\label{Table:Scenario2}
\centering
\setlength{\tabcolsep}{0.25em}
\def\arraystretch{1.3}
\begin{tabular}{|c|c|c|c|c|c|c|}
%\begin{tabular}{|c|c|S|S|S|S|S|S|}
\hline
\textbf{Beam} & \textbf{\begin{tabular}[c]{@{}c@{}}$\theta_{-3dB_o}$\\ ($^\circ$)\end{tabular}} & \multicolumn{1}{c|}{\textbf{\begin{tabular}[c]{@{}c@{}}$\theta_{-3dB_c}$\\ ($^\circ$)\end{tabular}}} & \multicolumn{1}{c|}{\textbf{\begin{tabular}[c]{@{}c@{}}error\\ (\%)\end{tabular}}} & \multicolumn{1}{c|}{\textbf{\begin{tabular}[c]{@{}c@{}}SLL\\ (dB)\end{tabular}}} & \multicolumn{1}{c|}{\textbf{\begin{tabular}[c]{@{}c@{}}Act.\\ Elem.\end{tabular}}} & \textbf{\begin{tabular}[c]{@{}l@{}}Directivity\\ (dBi)\end{tabular}}  \\ \hline
\rowcolor[HTML]{FFFFFF} 
1  & 4.7 & 4.7031 & 0.0663 & 16.813 & 118 & 28.166
\\ \hline
2  & 5.5  & 5.5041 & 0.0744 & 16.56 & 100 & 27.383 
\\ \hline                                                 
3  & 6 & 6.0245 & 0.4088 & 16.249  & 87 & 26.853  
\\ \hline
4 & 6.5 & 6.4523  & 0.7338 & 17.095 & 93 & 27.364
\\ \hline                            5 & 5.8 & 5.814 & 0.2413 & 16.766 & 95 & 27.366 
\\ \hline             
6 & 5 & 4.9863 & 0.2747 & 14.51 & 116 & 28.109        \\ \hline                       
7 & 7 & 7.0353 & 0.5042 & 14.422 & 73 & 26.283   
\\ \hline                                                                                                            
\end{tabular}
\end{table}
% Table3 end

\subsection{Scenario \#3}
In Scenario \#3, the beam synthesis process is subject to specific requirements, as outlined below:

\begin{itemize}
    \item \textbf{Activation instances:}  There is no limit in activation instances, which means that the algorithm will not control the activation but will control the achievement of the \ac{SLL} and beamwidth.
\end{itemize}

\begin{figure} [!htbp]  
\centering
\includegraphics[width=0.8\columnwidth]
{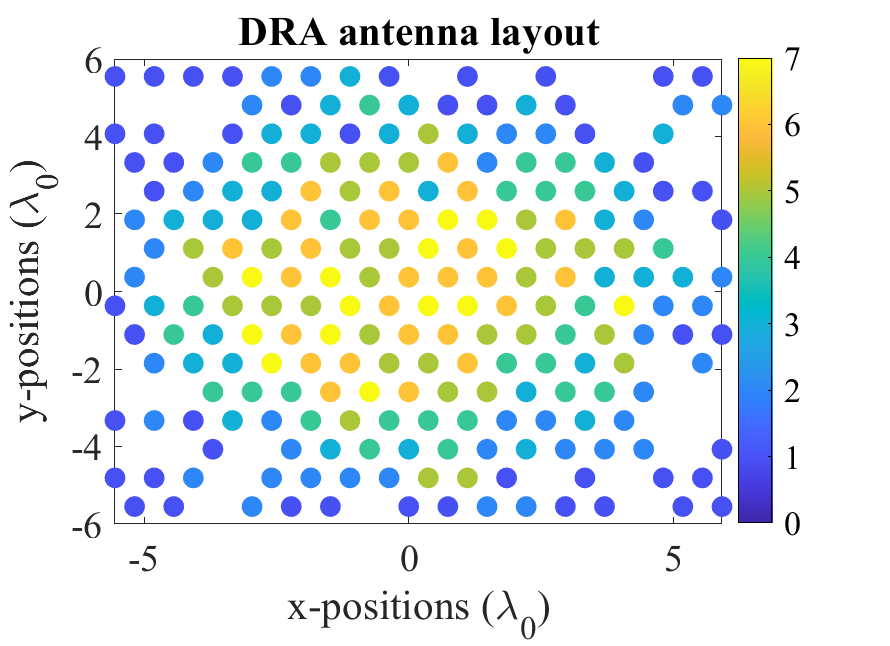}
\caption{Activation instances in the DRA for the Scenario \#3.}
\label{fig:Act7}
\end{figure}

Detailed results from the beam synthesis Scenario \#3 are presented in TABLE \ref{Table:Scenario3}. The activation instances for each antenna element in
Scenario \#3 is depicted in Fig. \ref{fig:Act7}

% Table 5
\begin{table}[!ht]
\caption{Characteristics of Synthesized Beams in the Scenario \#3.}
\label{Table:Scenario3}
\centering
\setlength{\tabcolsep}{0.25em}
\def\arraystretch{1.3}
\begin{tabular}{|c|c|c|c|c|c|c|}
%\begin{tabular}{|c|c|S|S|S|S|S|S|}
\hline
\textbf{Beam} & \textbf{\begin{tabular}[c] {@{}c@{}}$\theta_{-3dB_o}$ \\ ($^\circ$)\end{tabular}} & \multicolumn{1}{c|}
{\textbf{\begin{tabular}[c]
{@{}c@{}}$\theta_{-3dB_c}$ \\ ($^\circ$)\end{tabular}}} & \multicolumn{1}{c|}
{\textbf{\begin{tabular}[c]{@{}c@{}}error\\ (\%)\end{tabular}}} & \multicolumn{1}{c|}{\textbf{\begin{tabular}[c]{@{}c@{}}SLL\\ (dB)\end{tabular}}} & \multicolumn{1}{c|}{\textbf{\begin{tabular}[c]{@{}c@{}}Act.\\ Elem.\end{tabular}}} & \textbf{\begin{tabular}[c]{@{}l@{}}Directivity\\ (dBi)\end{tabular}}  
\\ \hline
\rowcolor[HTML]{FFFFFF} 
1  & 4.7 & 4.7003 & 0.0055 & 16.86 & 133 & 28.826
\\ \hline
2  & 5.5  & 5.4982 & 0.0322 & 17.259 & 111 & 27.979 
\\ \hline                                                 
3  & 6 & 6.0066 & 0.1096 & 16.824  & 102 & 27.824  
\\ \hline
4 & 6.5 & 6.4609  & 0.601 & 17.095 & 94 & 27.58
\\ \hline                           
5 & 5.8 & 5.8089 & 0.1534 & 16.766 & 104 & 27.837 
\\ \hline             
6 & 5 & 5.0025 & 0.0493 & 16.857 & 122 & 28.399       
\\ \hline                       
7 & 7 & 7.0434 & 0.6205 & 16.996 & 79 & 26.656   
\\ \hline                                                                                                            
\end{tabular}
\end{table}
% Table5 end

\subsection{Results Discussion}
According to results from each analyzed scenario, as the TABLES \ref{Table:Scenario1}, \ref{Table:Scenario2} and \ref{Table:Scenario3} depicts, the beamwidth variation and error fluctuation decreases as the number of the constrain of maximum activation instances increases. This pattern can be attributed
to the number of elements used for each beam, which limits the precision of beamwidth in all scenarios. The array pattern is calculated using the array multiplication method, which is a numerical method and is simply the product of the array factor and the radiation pattern of the unit cell in a particular direction \cite{vasquez2024multibeam}.

\begin{figure} [!htbp]  
        \centering
        \begin{subfigure} [b]{0.475\columnwidth}  
            \centering
            \includegraphics[width=\linewidth]{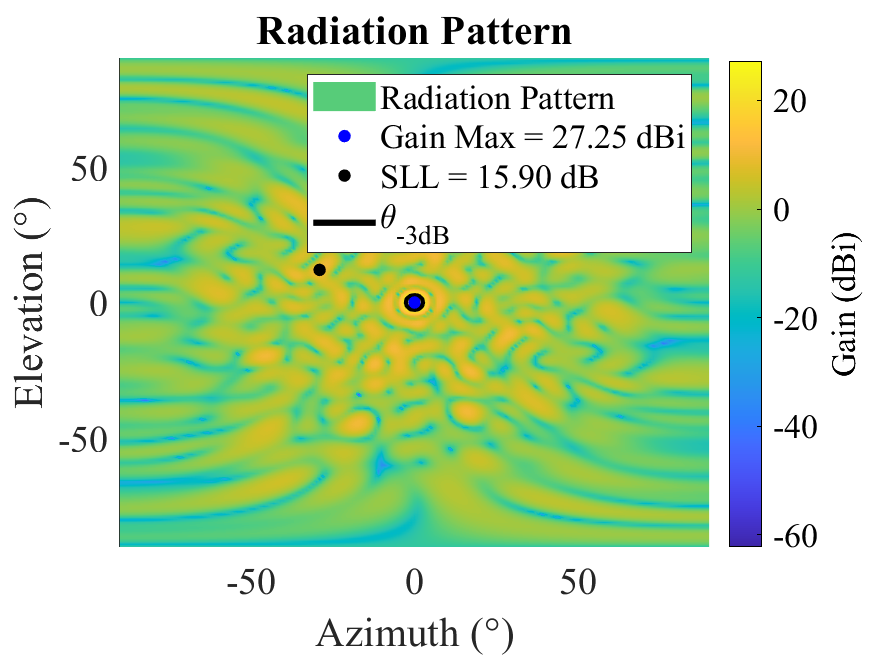}
            \caption[]%
            {{\small}}    
            \label{fig: B1}
        \end{subfigure}
        \hfill
        \begin{subfigure} [b]{0.475\columnwidth}  
            \centering 
            \includegraphics[width=\linewidth]{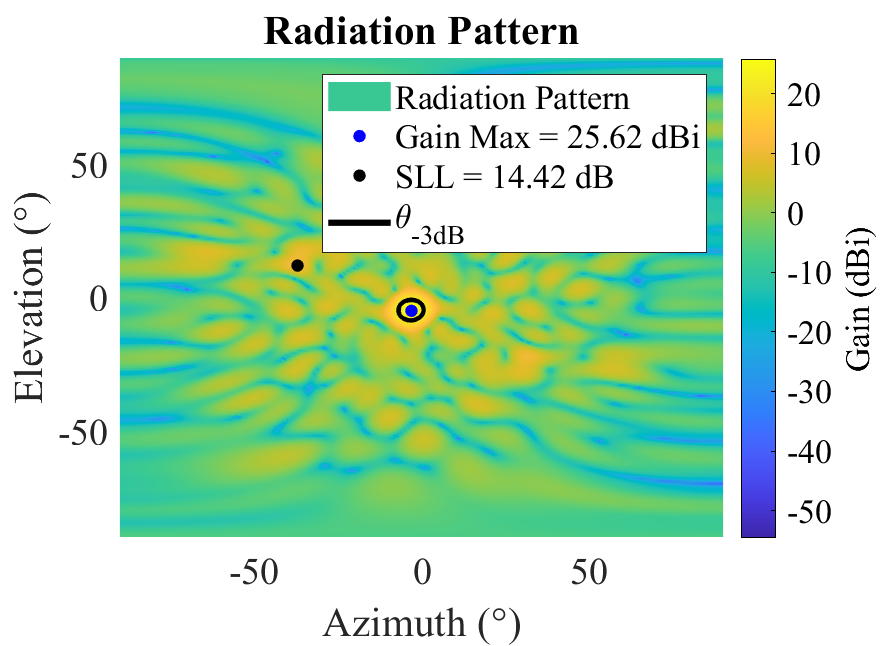}
            \caption[]%
            {{\small}}    
            \label{fig: B2}
        \end{subfigure}
        \vskip\baselineskip
        \begin{subfigure}  [b]{0.475\columnwidth}  
            \centering 
            \includegraphics[width=\linewidth]{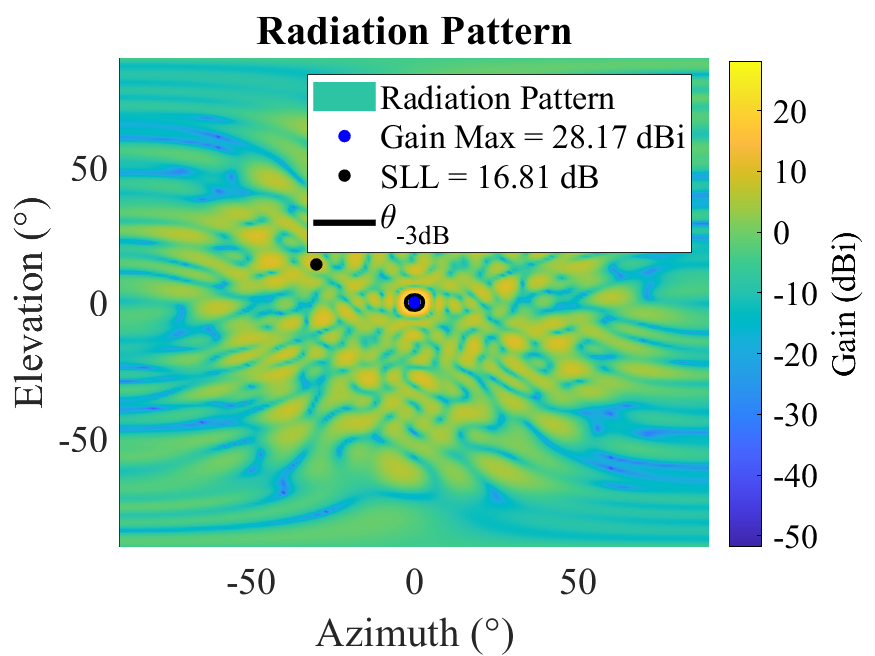}
            \caption[]%
            {{\small}}    
            \label{fig: B3}
        \end{subfigure}
        \hfill
        \begin{subfigure} [b]{0.475\columnwidth}  
            \centering 
            \includegraphics[width=\linewidth]{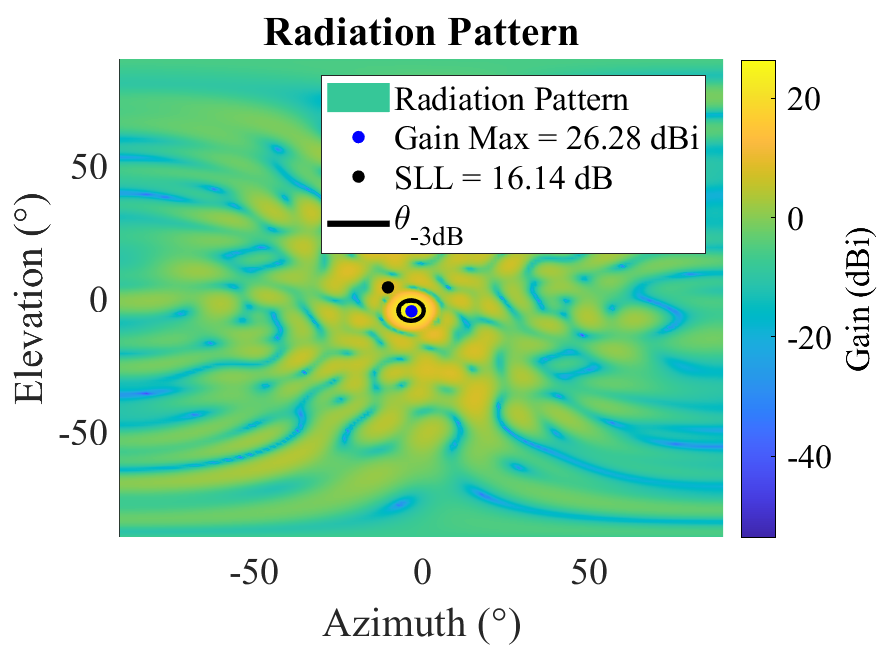}
            \caption[]%
            {{\small}}    
            \label{fig: B4}
        \end{subfigure}
        \vskip\baselineskip
        \begin{subfigure}  [b]{0.475\columnwidth}  
            \centering 
            \includegraphics[width=\linewidth]{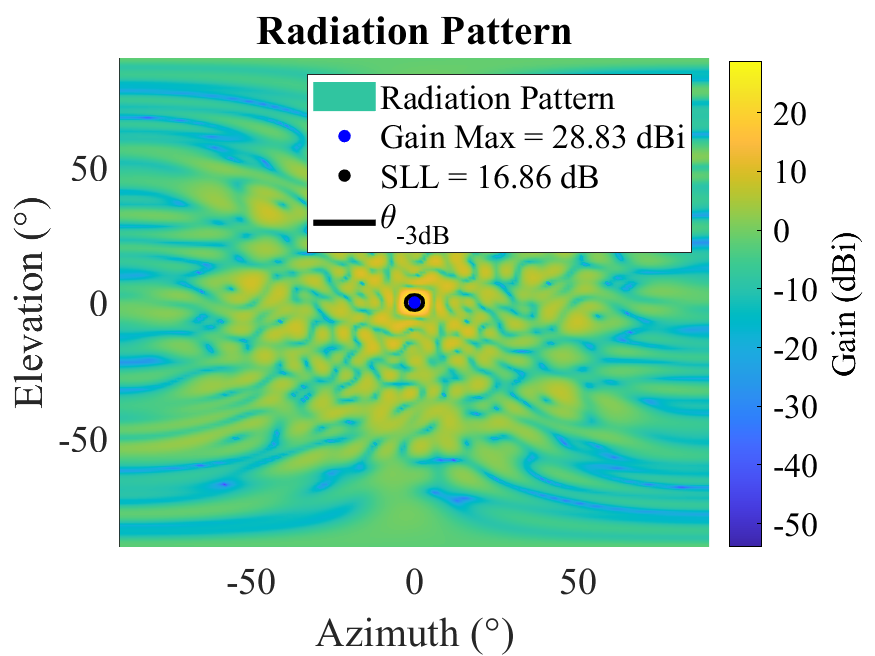}
            \caption[]%
            {{\small}}    
            \label{fig: B5}
        \end{subfigure}
        \hfill
        \begin{subfigure} [b]{0.475\columnwidth}  
            \centering 
            \includegraphics[width=\linewidth]{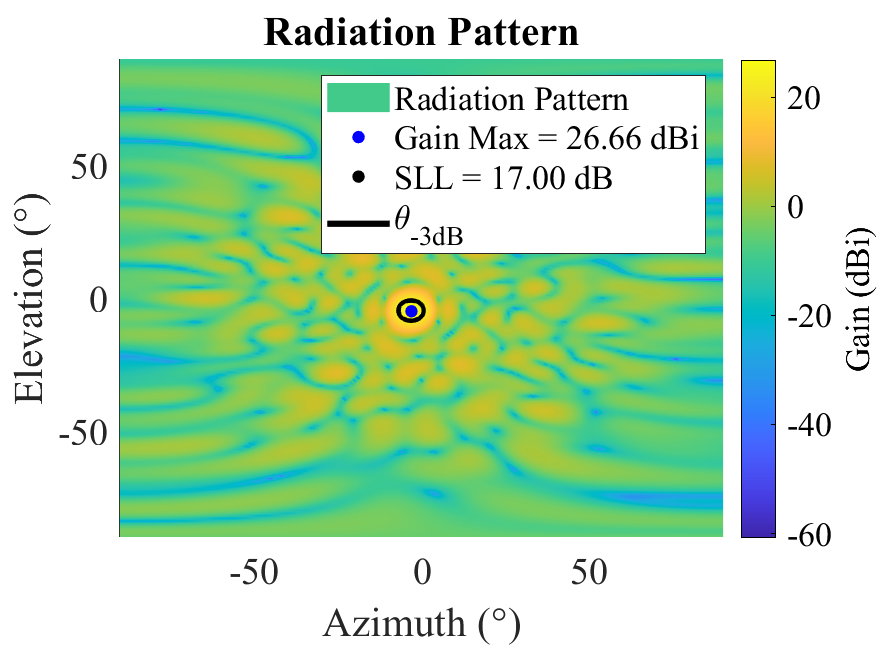}
            \caption[]%
            {{\small}}    
            \label{fig: B6}
        \end{subfigure}
        \caption[]
        {\small Radiation patterns in broadside of,  a) Beam \#1, Scenario \#1, b) Beam \#7, Scenario \#1, c) Beam \#1, Scenario \#2, d) Beam \#7, Scenario \#2, e) Beam \#1, Scenario \#3, and f) Beam \#7, Scenario \#3.} 
        \label{fig:RadPat}
    \end{figure}

On the other hand, it is observed that desired \ac{SLL} can not be achieved for all the beams in Scenario \#1 and Scenario \#2, where the activation instances are 5 and 6, respectively, for each antenna element. Since the number of elements required to form a beam reduces as the beamwidth grows there is a correlation between beamwidth and the number of activation instances. It should be noted that that tendency is more obvious in cases where the beamwidth approaches its lower limit. In contrast, in Scenario \#3, where the activation sample limit is not considered, it can be seen that all SLL values are below the desired level, regardless of the beamwidth. For a better inspection, the radiation patterns of the beams with the narrowest and widest beamwidths in scenarios Scenario \#1, \#2, and \#3 are given in \ref{fig:RadPat} in the broadside direction.    

As a final remark, TABLE \ref{Table:Scenario33} refers that the sum of the activated elements are almost the same, however, as the number of activation instances increases the overall directivity for each beam increases regardless of the beamwidth and the SLL. Moreover, Equivalent Isotropic Radiated Power (EIRP) per beam can be calculated as:
\begin{equation}
    \label{eq:EIRP}
    EIRP^b = G^b(\theta, \phi)(\text{dB}) + T_x (\text{dBW}), 
\end{equation}
where $G(\theta, \phi)$ is the gain of the optimized antenna in particular direction and $T_x$ is the transmitted power of the array.

% Table 6
\begin{table}[!ht]
\caption{Number of activation vs. Activated elements for Scenario \#1, Scenario \#2, and Scenario \#3.}
\label{Table:Scenario33}
\centering
\setlength{\tabcolsep}{0.25em}
\def\arraystretch{1.3}
\begin{tabular}{|c|c|c|c|}
%\begin{tabular}{|c|c|S|S|S|S|S|S|}
\hline
Number of & Scenario \#1 & Scenario \#2 & Scenario \#3  \\ Activation & Act. Ele. & Act. Ele. & Act. Ele. \\ \hline

1 & 46 & 41 & 45
\\ \hline
2 & 45 & 38 & 34 
\\ \hline                                                 
3 & 34 & 26 & 31 
\\ \hline
4 & 33 & 33 & 24
\\ \hline           
5 & 38 & 41 & 37
\\ \hline             
6 & 0 & 24 & 25     
\\ \hline                       
7 & 0 & 0 & 14  
\\ \hline                                                                                                            
\end{tabular}
\end{table}
% Table6 end

\section{Conclusion}
This study presents an approach for synthesizing beam patterns in multi-beam scenarios using a genetic algorithm. The algorithm takes into account parameters such as beamwidth, \ac{SLL}, element activation instances, and available power. Simulations and results demonstrate the capability of the proposed algorithm to distribute activation instances uniformly across multiple beams and generate effective beamforming. The proposed algorithm is able to prevent saturation in central antenna elements that are more prone to activation in multi-beam scenarios. 

Although the results are promising, this study can be extended by application to sector activation antenna instances, effective isotropic radiated power control, null steering (by applying one of the side-lobe cancellation techniques), and regular arrays. Such future research could improve the adaptability and efficiency of beamforming technology in a variety of communication contexts.

\bibliographystyle{IEEEtran}
\bibliography{References.bib}

\end{document}